\documentclass{elsart}

\usepackage{epsfig}

\begin{document}

\begin{frontmatter}

\title{Self-organized patterns of coexistence out of a
predator-prey cellular automaton}

\author{Kelly C. de Carvalho and }
\author{T\^{a}nia Tom\'{e}\corauthref{cor}}
\ead{ttome@if.usp.br}
\ead[url]{fig.if.usp.br/\~{}ttome}
\corauth[cor]{Corresponding author. Fax (55)-11-3813-4334.}

\address{
Instituto de F\'{\i}sica \\
Universidade de S\~{a}o Paulo \\
Caixa Postal 66318  \\
05315-970 S\~{a}o Paulo, SP, Brazil}

\date{\today}

\begin{abstract}

We present a stochastic approach to modeling the 
dynamics of coexistence of prey and predator populations.
It is assumed that the space of coexistence is explicitly subdivided in 
a grid of cells.
Each cell can be occupied by only one individual of each species or can be
empty. The system evolves in time according to
a probabilistic cellular automaton composed by a set of local rules
which describe interactions between species individuals 
and mimic the process of birth, death and  predation.
By performing computational simulations, we found that,
depending on the values of the parameters of the model,
the following states can be reached:
a prey absorbing state and active states of two types.
In one of them both species coexist in a stationary regime
with population densities constant in time. 
The other kind of active state is characterized by local 
coupled time oscillations of prey and predator populations.
We focus on the self-organized structures arising from
spatio-temporal dynamics of the coexistence.
We identify distinct spatial patterns of prey 
and predators and verify that they are intimally connected to the time 
coexistence behavior of the species.
The occurrence of a prey percolating cluster 
on the spatial patterns of the active states is also examined.

\end{abstract}


\end{frontmatter}

\section{Introduction}

The most renowned and perhaps the simplest model which displays
self-sustained coupled time oscillations in
a predator-prey system is the Lotka-Volterra model
(Lotka, 1920; Volterra, 1931). 
In this model individuals of each species
are dispersed over an assumed homogeneous space and their spatial
positions are not taken into account. It is implicitly considered
that any individual can interact with any other with
equal intensity and the time evolution of the species populations
is given by a set of two ordinary differential equations
(Lotka, 1920; Volterra, 1931).
These equations may be viewed
as mean-field type equations, which
do not take into account any spatial correlation
between individuals of each species.

If the descriptive level is one
where space structure is not relevant then a predator-prey model
constructed on the basis of a  mean-field approach, such as  
the Lotka-Volterra equations, can give
important qualitative information. It is possible yet 
to consider more complex prey-predator interactions in the
Lotka-Volterra model (Hastings, 1997) or more
sophisticated mean-field models 
(Satulovsky and Tom\'e, 1994; Satulovsky and Tom\'e, 1997;
Durrett and Levin, 2000; Ovaskainen et al., 2002; Aguiar et al., 2003),
which provide the stable coexistence of species and/or
stable population cycles. These improvements are
qualitatively relevant, since in the original
Lotka-Volterra model, the cycles
are not stable under small changes of the initial condition, which
is not biologically realistic (Hastings, 1997).

However, under certain ecological situations,
it is necessary to describe population dynamics by
using models which do take into account the spatial localization
and discreteness of individuals of each species. 
In fact, some experimental studies on predation,
as the one performed by
Huffaker (1958), show that
an inhomogeneous space is crucial for the maintenance of 
self-sustained time oscillations in a prey-predator system.
Therefore, to describe the oscillations and species coexistence,
it might be important  to consider a theoretical
approach that goes beyond the mean-field
type equations and is able to
incorporate the spatial structure of a system.
The r\^ole of space has actually been recognized in several
approaches for the description of different population biology
problems (Levin, 1974; Tainaka, 1988; Caswell and Etter, 1993; 
Durrett and Levin, 1994; Satulovsky and Tom\'e 1994;
Hastings, 1997; Hanski and Gilpin, 1997; Tilman and Kareiva, 1997; 
Provata et al., 1999; Liu et al., 2000; Antal and Droz, 2001;
King and Hastings, 2003; Aguiar et al., 2003; Carvalho and Tom\'e, 2004; 
Szab\'o and Sznaider, 2004; Stauffer et al., 2005; Nakagiri et al., 2005).

As summarized by Durrett and Levin (1994) there are
basically four theoretical approaches on population biology which
give descriptions at different levels. Among them we single out the
modeling by means of an interacting particle system 
(Liggett, 1985; Durrett, 1988; Marro and Dickman, 1999)
also known as a stochastic lattice model 
(Tainaka, 1988; Satulovsky and Tom\'e, 1994; Marro and Dickman, 1999;  
Antal and Droz, 2001; Tom\'e and de Oliveira, 2001)
in the context of nonequilibirum statistical mechanics. 
The main characteristic of this approach is that it
is based on spatial-structured models with continuous time
Markovian dynamics where
individuals are discrete and localized.

In the present work we are concerned with the modeling of a
predator-prey system by means of a
probabilistic cellular automaton where the individuals are discrete,
localized on the sites (or cells) of a lattice (or grid)
and interact only with their neighbors (local interactions). 
The system evolves in time according to a discrete
time stochastic Markovian  synchronous dynamics 
(Caswell and Etter, 1993; Tom\'e, 1994; Tom\'e and Drugowich 
de Fel\'{\i}cio, 1996).
In this approach the time is considered discrete and
the update is synchronous while maintaining the other
features of the description via an interacting particle system.
The local rules of the automaton mimic the process of predation, 
birth and death of individuals of the two species
and are inspired by the rules of the contact process 
(Liggett, 1985; Durrett, 1988; Marro and Dickman, 1999;
Tom\'e and de Oliveira, 2001).

In the following sections we present the model, explain the computational
simulation procedure, show the results,
and discuss the properties of the active states.
Our attention is concentrated on the study of the 
spatio-temporal dynamics of coexistence of species. 
We verify that a
given spatial distribution of prey and predators is
intimally connected to the time coexistence behavior of the species,
which may be oscillatory in time (in finite systems) or not.

\section{Model}

\subsection{Probabilistic cellular automata}

Our approach in the modeling of a predator-prey system by a 
probabilistic cellular automaton are based on the following assumptions. 
First, we consider that the space where the species interact
and survive is represented by a regular square lattice with $N$ sites.
Each one of the sites can be in one of three states:
empty (no individual species present), occupied by at most one
individual of the prey species or occupied by at most one
individual of the predator species.
These states will be represented by a random variable 
$\eta _{i}$, associated to site
$i$, which takes three values $0$, $1$ or $2$, according whether the site
$i$ is empty, occupied by one prey individual or occupied by 
one predator individual. 

\begin{figure}
\centering
\epsfig{file=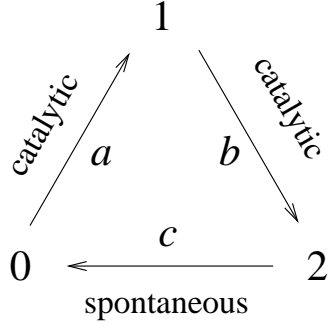,width=4.5cm}
\caption{A scheme of the allowed transitions between distinct states
showing the type of process and respective probabilities.
The numbers 0, 1, and 2 represent an empty site, a
site occupied by a prey individual and a
site occupied by a predator individual, respectively.}
\label{scheme}
\end{figure}

Each site can assume one state at each
time step and the transitions between the possible
states, which occur in successive instants of time, obey stochastic
local rules that define the
prey-predator interaction in a local level. We assume that the
transitions from one state to another distinct state
must obey the clockwise cyclic ordering shown in Figure \ref{scheme}.
The counterclockwise cyclic ordering is forbidden.
This assumption implies that the model is microscopically
irreversible or that it lacks detailed balance 
(van Kampen, 1981; Tom\'e and de Oliveira, 2001).
Prey can proliferate just on empty sites; prey
give place to predator in the predation process, where it is
implicit that a predator reproduces and the prey dies; and a
predator can die leaving an empty site. The empty sites can be seen
as the resource of food for prey surveillance and proliferation. The
process of death of predator complete this cycle reintegrating these
resources to the system. Two of the transition rules are catalytic:
the birth of prey in a empty site is conditioned to the existence of
prey in its neighborhood on the lattice; and the birth of
predator in a site occupied by prey is conditioned to the existence
of predators in its neighborhood. The process of death of
predator is spontaneous, that is, it occurs independently of the
states of the neighbors. The conditions for
the survival of the species are the same in any region of the
space implying that the rules are the same for any site of the
lattice. The three processes, the birth of prey, birth of predator 
and death of predator, comprehend, at a microscopic level, 
the same reactions involved in the Lotka-Volterra model 
and are associated to the parameters $a$, $b$, and $c$, respectively.

Let us denote by 
$P_{\ell }(\eta )$, 
the probability of state 
$\eta =(\eta _{1},\eta_{2},...,\eta _{N})$ 
of the system, at time $\ell $. As we will consider a
Markovian discrete time and discrete space process the evolution equation for
the probability is
\begin{equation}
P_{\ell +1}(\eta )=\sum_{\eta ^{\prime }}
T(\eta | \eta ^{\prime})P_{\ell }(\eta ^{\prime }),
\label{mas}
\end{equation}
where the sum is over all possible states of the system, and 
$T(\eta | \eta ^{\prime })$ is the conditional transition 
probability from a state $\eta^{\prime }$ to state $\eta $, 
given that at the previous time step the
system was in state $\eta ^{\prime }$. 
Since we would like to model the system using a cellular automaton, 
the update of the sites is synchronous.
The global transition probability $T(\eta | \eta ^{\prime })$
is written as the product of the transitions probabilities for each site
\begin{equation}
T(\eta | \eta ^{\prime })=
\prod_{i=1}^{N} w_{i}(\eta _i |  \eta^\prime), 
\label{matrix_product}
\end{equation}
where we have denoted by $w_{i}(\eta _i |  \eta^\prime)$ 
the probability that site $i$ assumes the state 
$\eta _{i}$ given that the system is in state $\eta ^{\prime }$
at the previous time. We observe that $T(\eta|\eta^\prime)$ 
can be written as a product because the
state assumed by each site in a given time step $\ell $ is independent of
the states assumed by the other sites. 
We also remark that from the
property that the distribution probability is normalized,
the following  properties must be held
\begin{equation}
w_{i}(\eta _{i} | \eta ^{\prime })\geq 0
\qquad\qquad
{\rm and}
\qquad\qquad
\sum_{\eta _{i}} w_{i}(\eta _{i} | \eta^\prime)=1.
\label{property_w}
\end{equation}
A probabilistic cellular automaton is then defined by just giving the
set of transition probabilities, or local rules, 
$w_{i}(\eta _{i} | \eta ^{\prime })$. 

\subsection{Predator-prey probabilistic cellular automaton}

Here we describe the probabilistic 
cellular automaton for the predator-prey system.
Some aspects of this automaton has been considered
in a preliminary study by Carvalho and Tom\'e (2004).
A site $i$ of a
regular square lattice is updated at each time step by considering
its interaction with its neighborhood defined as the 
four nearest neighbor sites at north, east, west and south. 
The lattice is synchronously updated and each site in the lattice
changes its state according to the set of transition probabilities,
or local rules, $w_i(\eta _{i}^{\prime }|\eta )$ defined as follows.

{\it Birth of prey}. 
If a site $i$ is empty, $\eta _{i}=0$, then it can be occupied
by a prey individual if there are individuals of the same species
in its neighborhood. The probability of the transition
$0\to1$ is equal to $(a/4)$ times the number $N_{1,i}$
of prey individuals present in the four sites of the neighborhood. No birth
occurs in the absence of prey, which means that this 
is a catalytic process. 

{\it Predation and birth of predator}.
The predation occurs when a given site $i$
is occupied by prey, $\eta_{i}=1$,
and there are predators in its neighborhood. 
At the same time a predator is born at site $i$.
The probability of the transition $1\to2$
is equal to $(b/4)$ times the number $N_{2,i}$
of predator individuals in the four sites of the neighborhood.
No death of prey and simultaneous birth of predator 
occur in the absence of predators, which means that this 
is also a catalytic process.

{\it Death of predator}.
The death of predator occurs spontaneously,
that is, it does not depend on the
state of the neighboring sites. 
In this process a site $i$ which is
occupied by a predator, $\eta _{i}=2$, is evacuated with probability
$c$, that is, the probability of the transition $2\to0$ is equal to $c$.

Formally we may write the transition probabilities 
corresponding to a generic site $i$ as follows
\begin{equation}
w_i(1|\eta) = \frac{a}{4}N_{1,i}\, \frac{(\eta_i-1)(\eta_i-2)}{2}
+(1-\frac{b}{4}N_{2,i})\, [\eta_1(2-\eta_i)],
\label{4}
\end{equation}
\begin{equation}
w_i(2|\eta) = \frac{b}{4}N_{2,i}\,  [\eta_i(2-\eta_i)]
+ (1-c)\, \frac{\eta_i (\eta_i-1)}{2},
\label{5}
\end{equation}
and
\begin{equation}
w_i(0|\eta) = c\, \frac{\eta_i (\eta_i-1)}{2}
+(1-\frac{a}{4}N_{1,i})\, \frac{(\eta_i-1)(\eta_i-2)}{2}.
\label{6}
\end{equation}
They are better viewed in the following matrix
\begin{equation}
\left(
\begin{tabular}{ccc}
$1-aN_{1,i}/4$  &   0              &  $c$   \\
$aN_{1,i}/4$  &   $1-bN_{2,i}/4$   &  0              \\
0             &   $bN_{2,i}/4$   &  $1-c$   \\
\end{tabular}
\right)
\end{equation}
where each entry of the matrix represents
the transition probability from the state defined
by the column index (0, 1, or 2) to the state defined by the row index
(0, 1, or 2).

These local rules represent the interactions of
a system of particles on a lattice. They
have some similarities with the local rules  of
the contact process 
(Liggett, 1985; Durrett, 1988; Marro and Dickman, 1999;
Tom\'e and de Oliveira, 2001),
although here we have
a three state discrete time Markovian process whereas
the contact process is a two state Markovian continuous time process.
The dynamics defined by the above rules 
leads to active states in which prey and predator coexist 
and to absorbing states in which
the system become trapped.  

{\it Empty absorbing state}.
If a state devoid of prey is reached, there will be no
birth of predator and, since 
the death of predators is spontaneous, a state devoid
of individuals of any species will be attained. The system is then trapped
in this empty state where both  species have been extinct.
This state, however, does not occur due to the abundant
population of prey.

{\it Prey absorbing state}.
If a state devoid of predator is reached 
and prey have not been extincted then, prey
will reproduce until they will cover the entire lattice.
The system is then trapped in this prey absorbing state.

The presence of the absorbing states implies the lack of
detailed balance which means that the
model is intrinsically irreversible 
(van Kampen, 1981; Marro and Dickman, 1999; Tom\'e and de Oliveira, 2001).

\begin{figure}
\centering
\epsfig{file=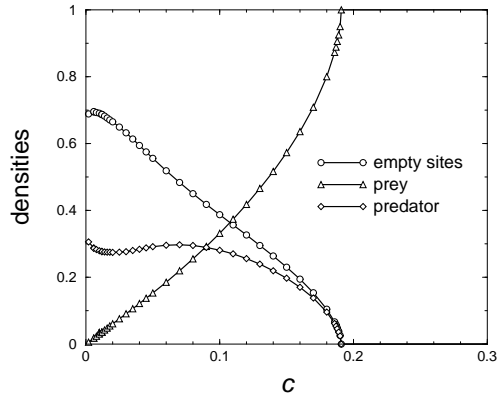,width=6.5cm}
\caption{Densities of prey, predator and empty sites versus $c$
for $p=0.$}
\label{dens}
\end{figure}

\section{Simulation of the cellular automaton}

\subsection{Active states}

In order to analyze the behavior of the predator-prey system 
modeled by the cellular automaton we perform numerical simulations.
We consider a square lattice of $N=L^{2}$ sites and use
periodic boundary conditions, that is, opposite edges of the lattice
are connected forming a torus. The initial configuration
is randomly generated by placing prey with probability 1/3,
predators with probability 1/3 and leaving the sites
empty with probability 1/3. 
Each site is updated, synchronously and independently,
according to the rules
(\ref{4}), (\ref{5}) and (\ref{6}).
Most of our results were obtained for lattice sizes with
$N=160\times 160$ sites. 

Each update of the lattice corresponds to one time
step. The first steps in the simulation are discarded, since they correspond
to the transient regime where microscopic configurations are 
not yet being generated according to the appropriate 
probabilities specified by the given set of parameters $a$, $b$ and $c$.
The density of prey $\rho_{1}$, the density of predator $\rho _{2}$
and the density of empty sites $\rho_0=1-\rho_1-\rho_2$  
are computed for each time step.

\begin{figure}
\centering
\epsfig{file=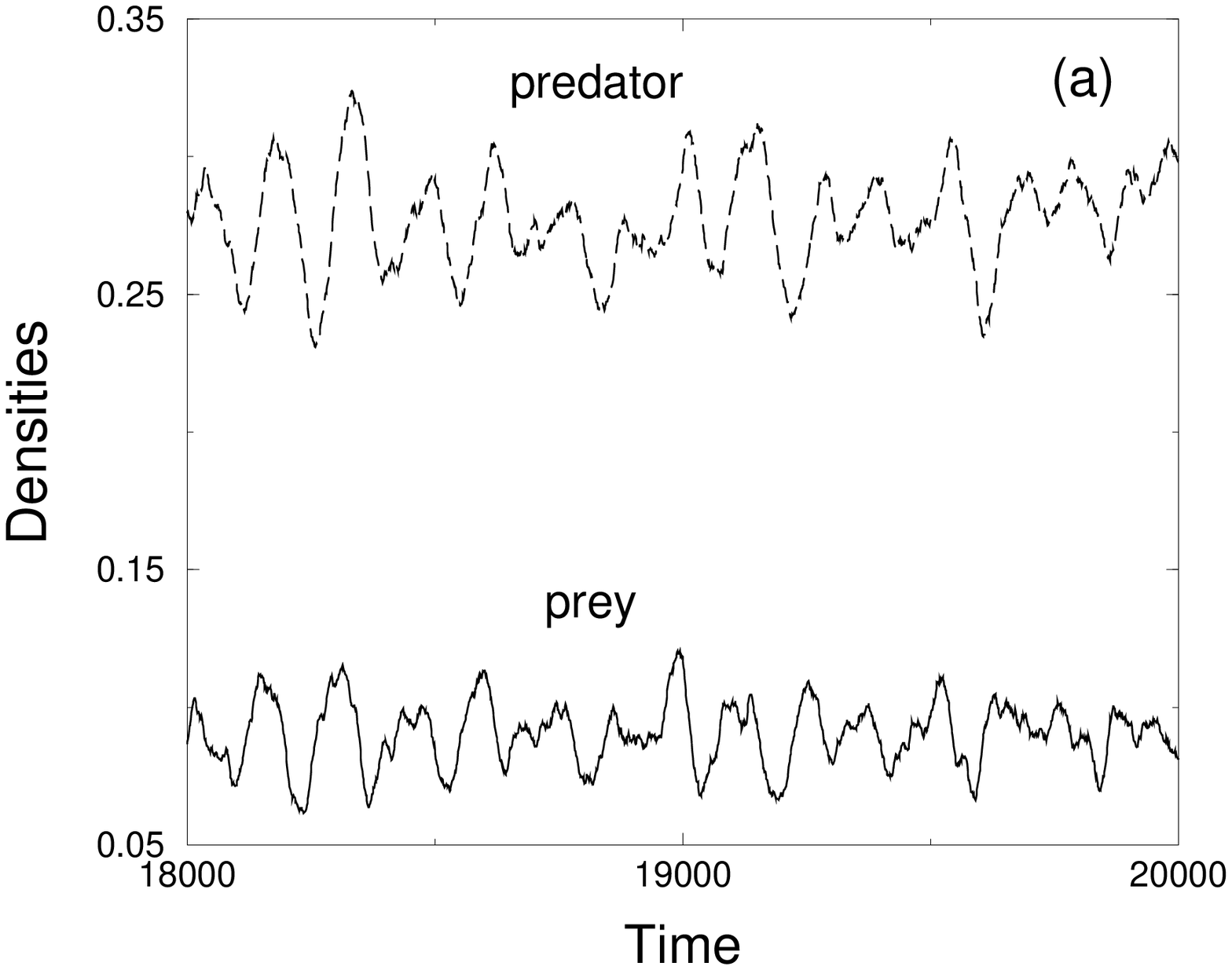,width=6.5cm}
\hfill
\epsfig{file=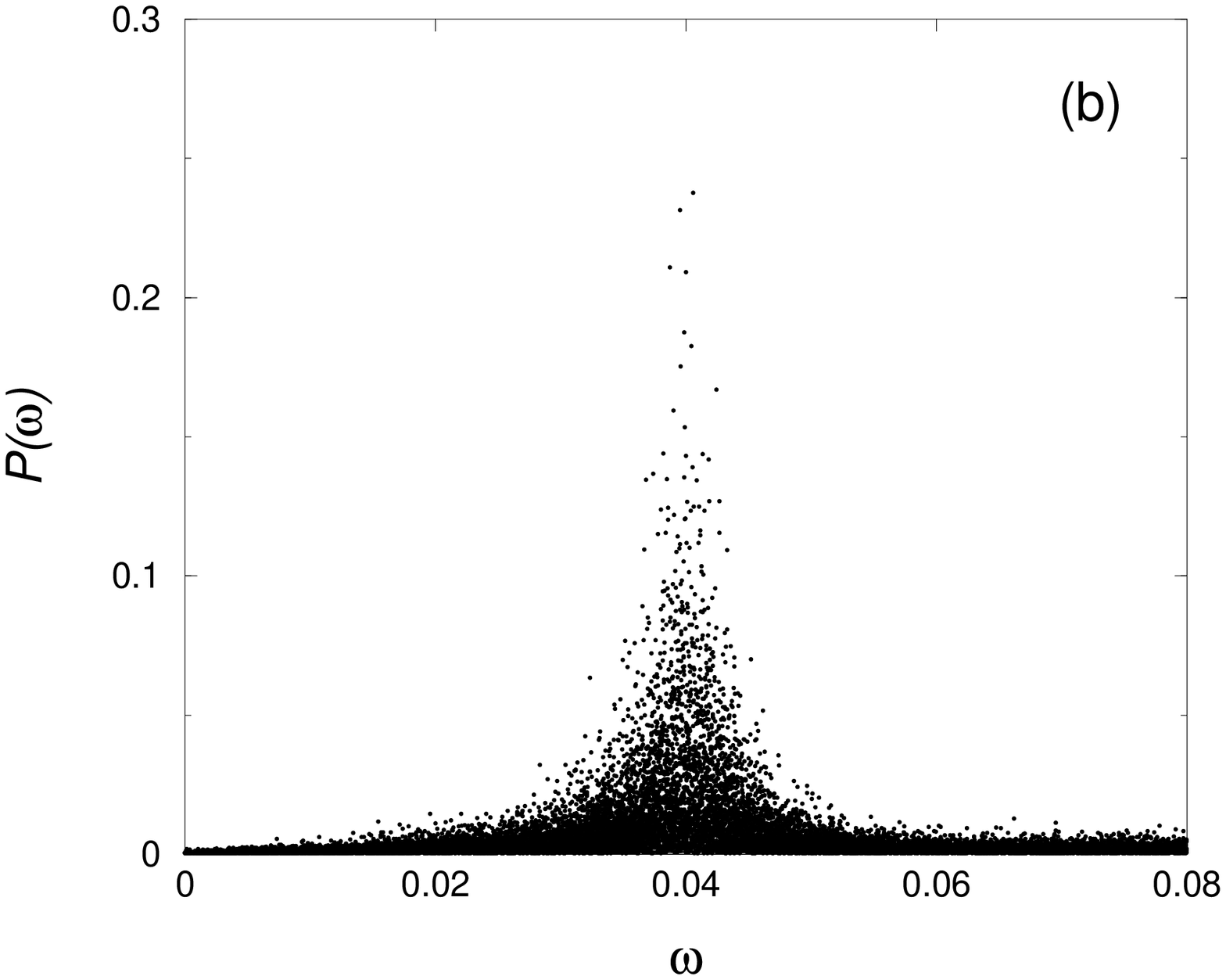,width=6.5cm}
\caption{(a) Densities of prey and predator as a function of 
time for $p=0$ and $c=0.03$ (oscillating state). 
(b) Corresponding power spectrum for the
density of prey as a function of the frequency $\omega$.
Results obtained for a lattice of size $160\times160$.}
\label{osc}
\end{figure}

\begin{figure}
\centering
\epsfig{file=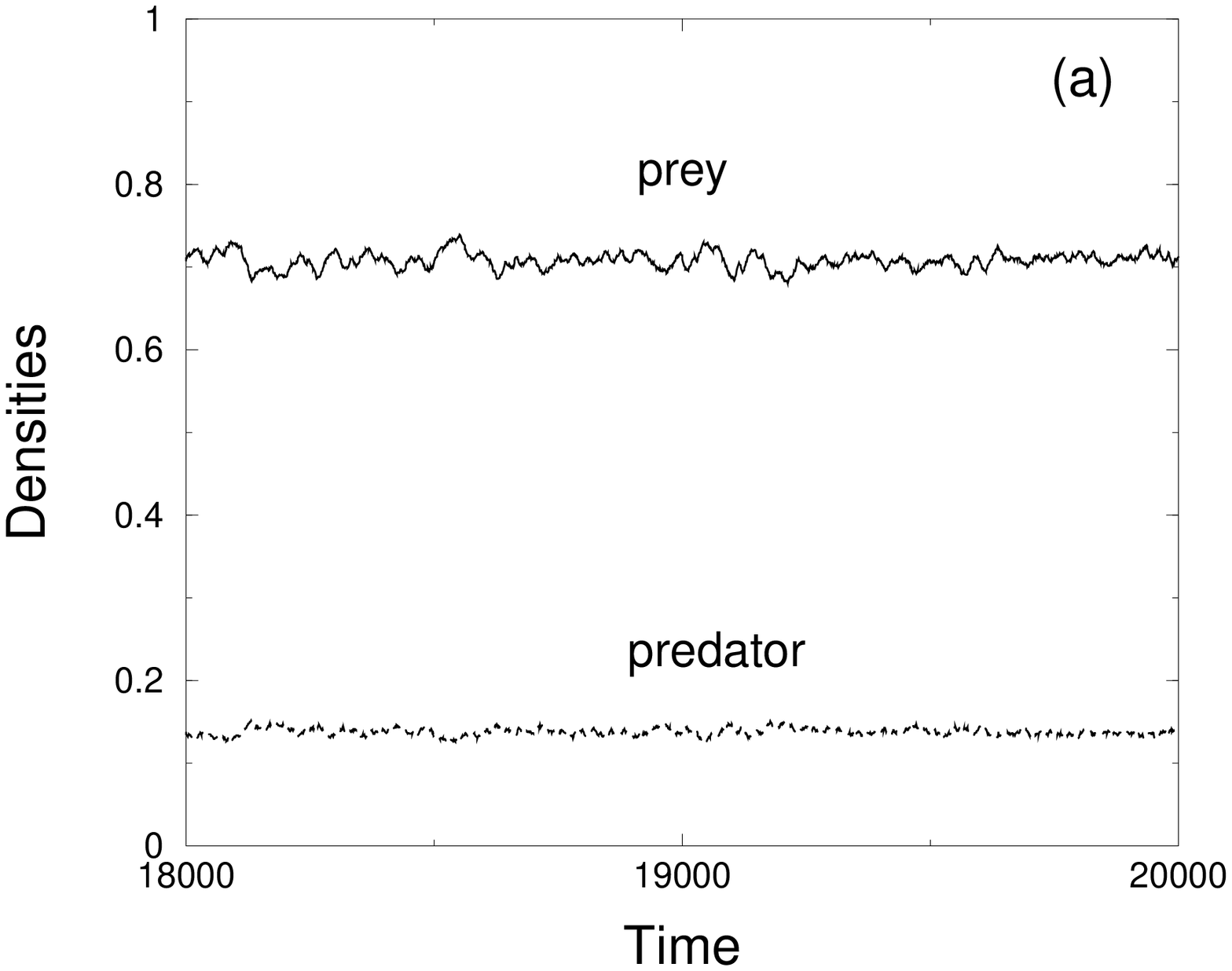,width=6.5cm}
\hfill
\epsfig{file=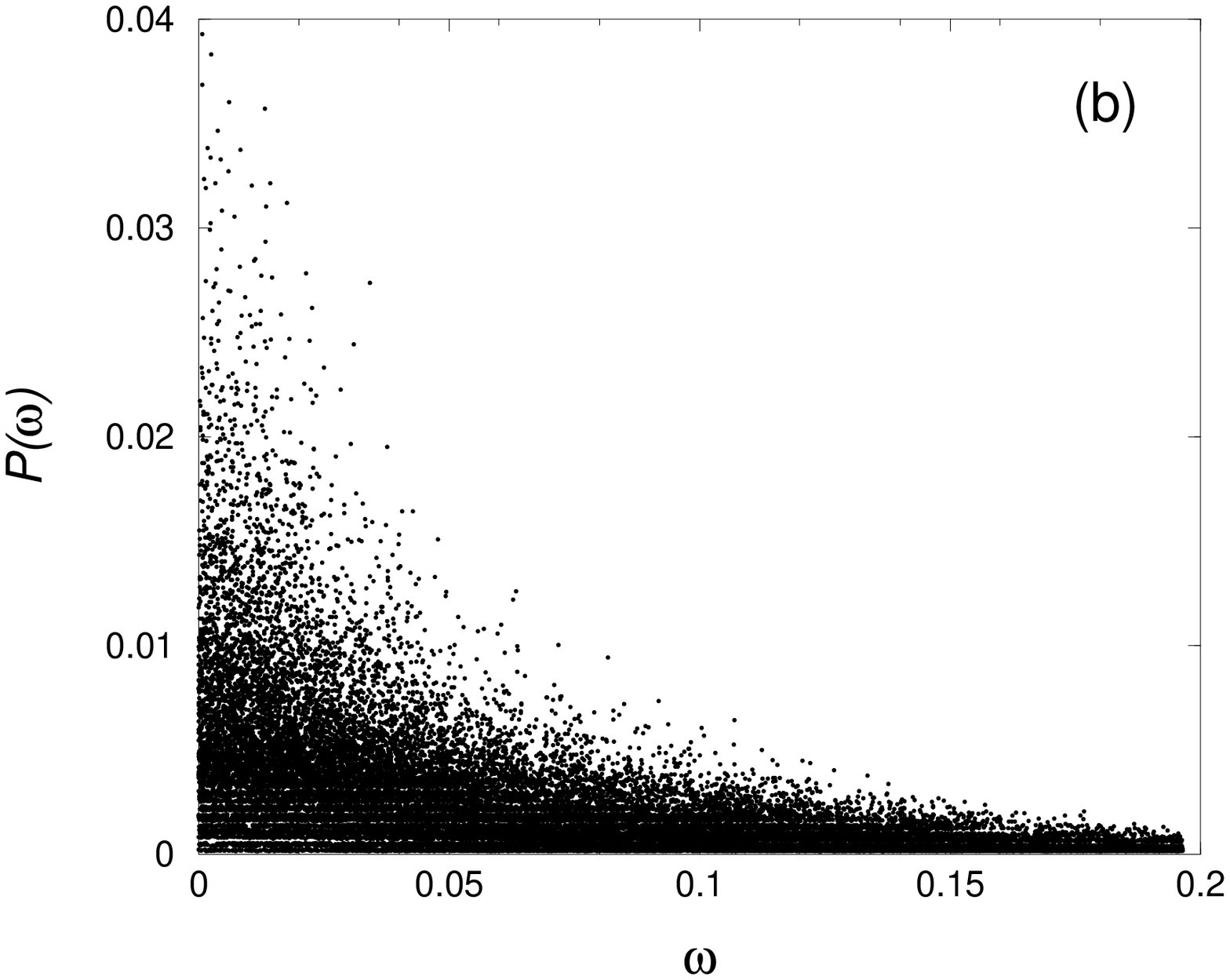,width=6.5cm}
\caption{(a) Densities of prey and predator as a function of 
time for $p=0$ and $c=0.17$ (nonoscillating state). 
(b) Corresponding power spectrum for the
density of prey as a function of the frequency $\omega$.
Results obtained for a lattice of size $160\times160$.}
\label{nosc}
\end{figure}

We restrict ourselves to the values of $a$, $b$ and $c$
such that $a+b+c=1$. 
This allows us to introduce the following parametrization
\begin{equation}
a=({1-c})/{2}-p
\qquad\qquad
{\rm and}
\qquad\qquad
b=({1-c})/{2}+p,
\label{param}
\end{equation}
with $-1/2\leq p\leq 1/2$. The parameter $c$, the death
of predators, is restricted to 
$0\leq c\leq 1-2\left\vert p\right\vert $.

Fixing the value of $p$ we have found that the system evolves in
time and eventually reaches either an absorbing prey state or an active
state. An absorbing prey state, which is characterized by $\rho
_{1}=1,\rho _{2}=\rho _{0}=0$, occurs for high values of $c.$ As $c$
is decreased the active state, where $0<\rho _{1}<1$, $0<\rho
_{2}<1,0<\rho _{0}<1$, is reached. The behavior of the densities of
prey, predators and empty sites versus $c$, for $p=0$, is shown in
Figure \ref{dens}.

Figures \ref{osc} and \ref{nosc} exhibit two possible time evolutions
of the densities of prey and predators corresponding to active states.
For small values of $c$ the time series for the population
densities show an oscillatory behavior,
as can be seen in Figure \ref{osc}. 
For larger values of $c$ the oscillations disappear and
the observed density variation in this time series are just stochastic
fluctuations, as shown in Figure \ref{nosc}.
Typical power spectra $P(\omega)$ related to the density of prey,
for the oscillating and nonoscillating cases, 
are also shown in Figures \ref{osc} and \ref{nosc}, respectively.
The power spectra related to the density of predator
(not shown) are similar.
The presence of a prominent and
well defined peak in the power spectrum of Figure \ref{osc}
at a nonzero frequency
characterizes an oscillating behavior. 
The power spectra
for the population densities of predators and prey have a peak at
the same frequency implying that the oscillations of the two
species are coupled. In contrast, the monotonic decreasing in the
power spectrum with frequency characterizes a random fluctuation
related to the nonoscillating active state, as shown in 
Figure \ref{nosc}.

\begin{figure}
\centering
\epsfig{file=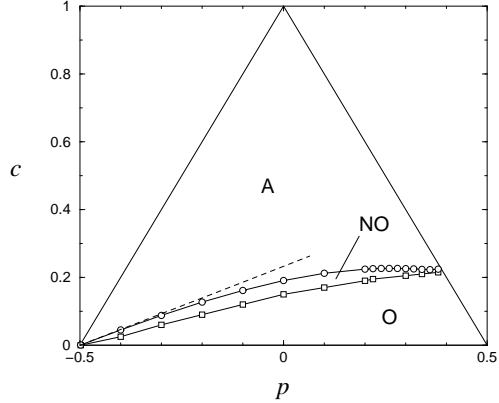, width=6.5cm}
\caption{Space of parameters, or $p-c$ diagram,
showing regions corresponding to three states: 
prey absorbing (A), active nonoscillating (NO) 
and oscillating (O). The straight dashed line
represents the critical transition line of the 
contact process for the predator-prey system
without vacant sites (see text).}
\label{diag}
\end{figure}

\subsection{Space of parameters}

The parametrization defined by equation (\ref{param}), which implies the 
condition $0\leq c \leq 1-2|p|$,  means that the possible
values of $c$ and $p$ defining the space of parameters,
or the $p-c$ diagram,
are restricted to the triangle drawned in Figure \ref{diag}.
Depending on the values of the parameters $c$ and $p$
one of the three possible states, prey absorbing and the two active states,
can be reached as shown in Figure \ref{diag}.
This diagram was obtained by numerical simulation with lattice sizes
up to $N=720\times 720$ sites.
A critical transition line $c_1(p)$ from
the prey absorbing state to the active state
cross the entire $p-c$ diagram starting from the left
corner of the triangle and ending at the opposite side.
The active region of the $p-c$ diagram is divided into two regions,
the oscillating and nonoscillating, by a line $c_2(p)$ that
also starts at the left corner and ends at the 
opposite side of the triangle.

The presence of a peak at a nonzero frequency in the
power spectrum was used to estimate the transition line 
on the $p-c$ diagram between the regions corresponding to the
oscillating active state and the non-oscillating active state.
Fixing the value of $p$ and departing from the oscillating region
of the diagram we increase the value of $c$. 
The transition to the nonoscillating active region occurs at a value $c_2$
where the peak in the power spectrum disappears.
For instance, for $p=0$, we get $c_2 \approx 0.15$ at which 
point the prey density assumes the value $\rho_1 \approx 0.58$.
It is worth to note that the oscillatory behavior is observed
in finite systems what implies that the oscillations occur
in a local level (Carvalho and Tom\'e, 2004).

To analyze the transition between the regions of the diagram
corresponding to the active state
and the prey absorbing state we have considered large lattice sizes.
Also, we have modified the simulation procedure by not allowing the
extinction of the species: if the number of prey vanishes then 
a prey individual is created in an empty
site chosen at random; if the number of predators
becomes zero then we randomly choose a site occupied
by prey and replaces it with a predator. This makeshift
was required mainly in the transient regime where 
greater amplitudes and fluctuations are attained.
A spurious entrance in the absorbing state was 
avoided by the makeshift and the conditions for
a stable regime were provided. 

\begin{figure}
\centering
\epsfig{file=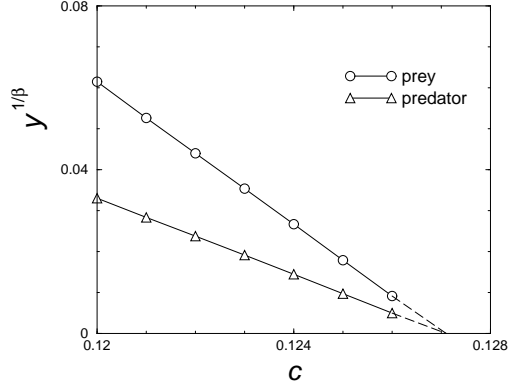,width=6.5cm}
\caption{
The quantity $y^{1/\beta}$, where $y$ represents either
the predator density $\rho_2$ or $1-\rho_1$, where $\rho_1$ is the
prey density, versus $c$, with 
$\beta=0.58$ (Marro and Dickman, 1999).
Results correspond to $p=-0.2$.}
\label{colapso}
\end{figure}

We expect that near the transition from the active state
to the prey absorbing state the
densities of predator and prey obey the following 
asymptotic behavior
$\rho_2 \sim (c_{1}-c)^{\beta}$
and
$1-\rho_1 \sim (c_{1}-c)^{\beta}$,
respectively, where $\beta$ is the critical exponent
associated with the order parameter (Marro and Dickman, 1999).
From Figure \ref{colapso} we can see that, fixing $p$, the densities of 
predators and prey indeed obey these relations.
With this assumption we were
able to estimate the localization of the critical line $c_1(p)$
from the active state to
the absorbing state by fitting a straight line to the
data points of $(1-\rho_1)^{1/\beta}$ 
and $\rho_2^{1/\beta}$ versus $c$.
It was assumed that the critical exponents associated to the
transition from the active state to the absorbing state 
belong in the direct percolation universality class
(Grassberger, 1982),
which gives a critical exponent $\beta=0.58$
in two dimensions (Marro and Dickman, 1999).

\subsection{Contact process limit}

\begin{figure}
\centering
\epsfig{file=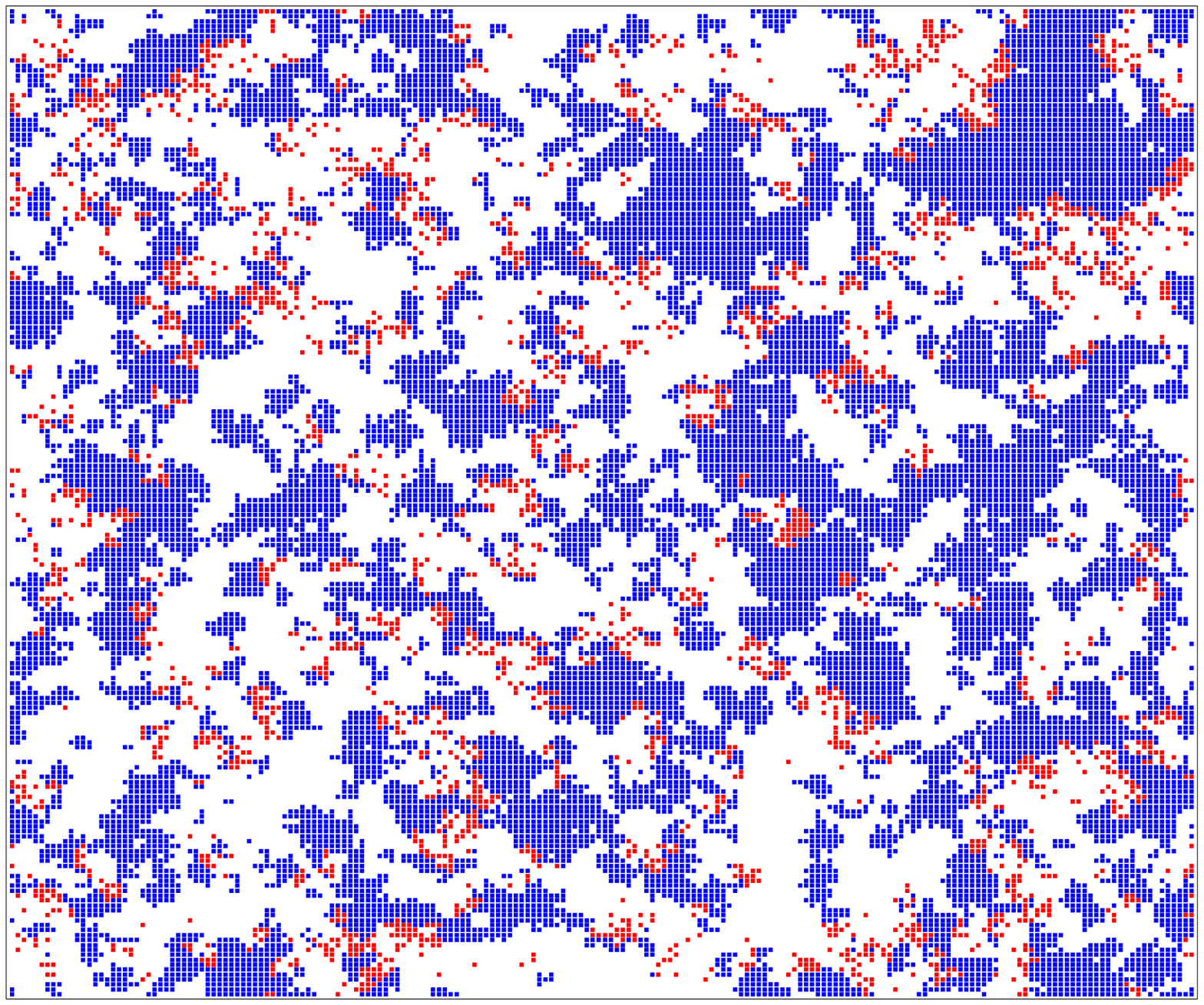,angle=-90,width=6.5cm}
\epsfig{file=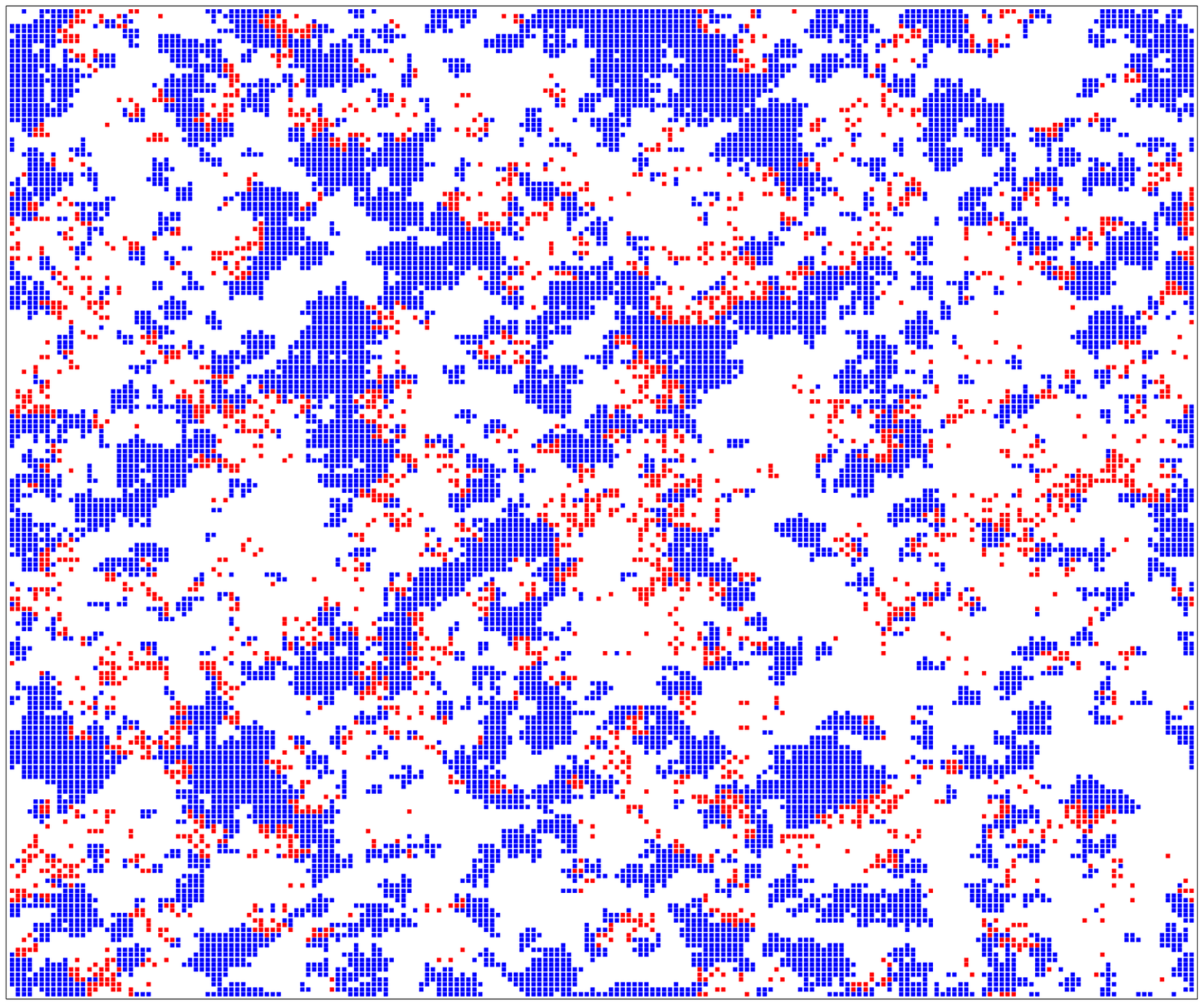,angle=-90,width=6.5cm}
\caption{Snapshots of the lattice for $p=0.3$ and $c=0.13$
(inside the oscillating region of the $p-c$ diagram)
taken at a maximum of prey (left) and a minimum of prey (right).
Results obtained for a lattice of size $160\times 160$.
The blue points represent sites occupied by prey, the red points 
by predators and the white points are empty sites.}
\label{mancha}
\end{figure}

As the probabilities of death and birth of predators become negligible 
($c$ and $b$ small, corresponding to the region around the left 
corner of the $p-c$ diagram of Figure \ref{diag}), the dynamics of the model
is dominated by the birth of prey. It means that any empty site is quickly
converted into a site occupied by prey. Since the spontaneous
death of predator create empty sites and these in turn are
almost instantaneously occupied by prey,
one may say that
predators are being spontaneously ``converted'' into prey.
We may thus replace the two
reactions $2\to 0$ and $0\to 1$ by just a spontaneous
reaction $2\to 1$. The whole process is 
therefore reduced to a contact process with two states (1 and 2, or predator
and prey) with a spontaneous reaction $2\to 1$ with probability $c$
and a catalytic reaction $1\to 2$ with probability $b$.
Equivalently, the spontaneous process
occurs with rate 1 and the catalytic process with creation rate 
$\lambda=b/c$. For $\lambda < \lambda_c$, where $\lambda_c$ is the
critical creation rate, the stationary
state is the absorbing state. 
Then, for $b<\lambda_c \,c$
the system displays the prey absorbing state. For small 
values of $c$ the critical line is given by $b=\lambda_c \,c$
or, equivalently, $c= (p+0.5)/(\lambda_c+0.5)$.
Our numerical simulations confirm this conjecture if we use 
the value $\lambda_c=1.65$ for the contact process
in a square lattice (Marro and Dickman, 1999).
This result can be seen on the diagram of 
Figure \ref{diag} where we have plotted the line 
$c= (p+0.5)/(\lambda_c+0.5)$. This line is tangent to
the critical transition line $c_1(p)$
at the left corner point of the triangle.

\section{Spatial patterns and coexistence of species}

There are basically three types of spatial pattern formation
coupled to the coexistence of species. They are described
as follows.

\begin{figure}
\centering
\epsfig{file=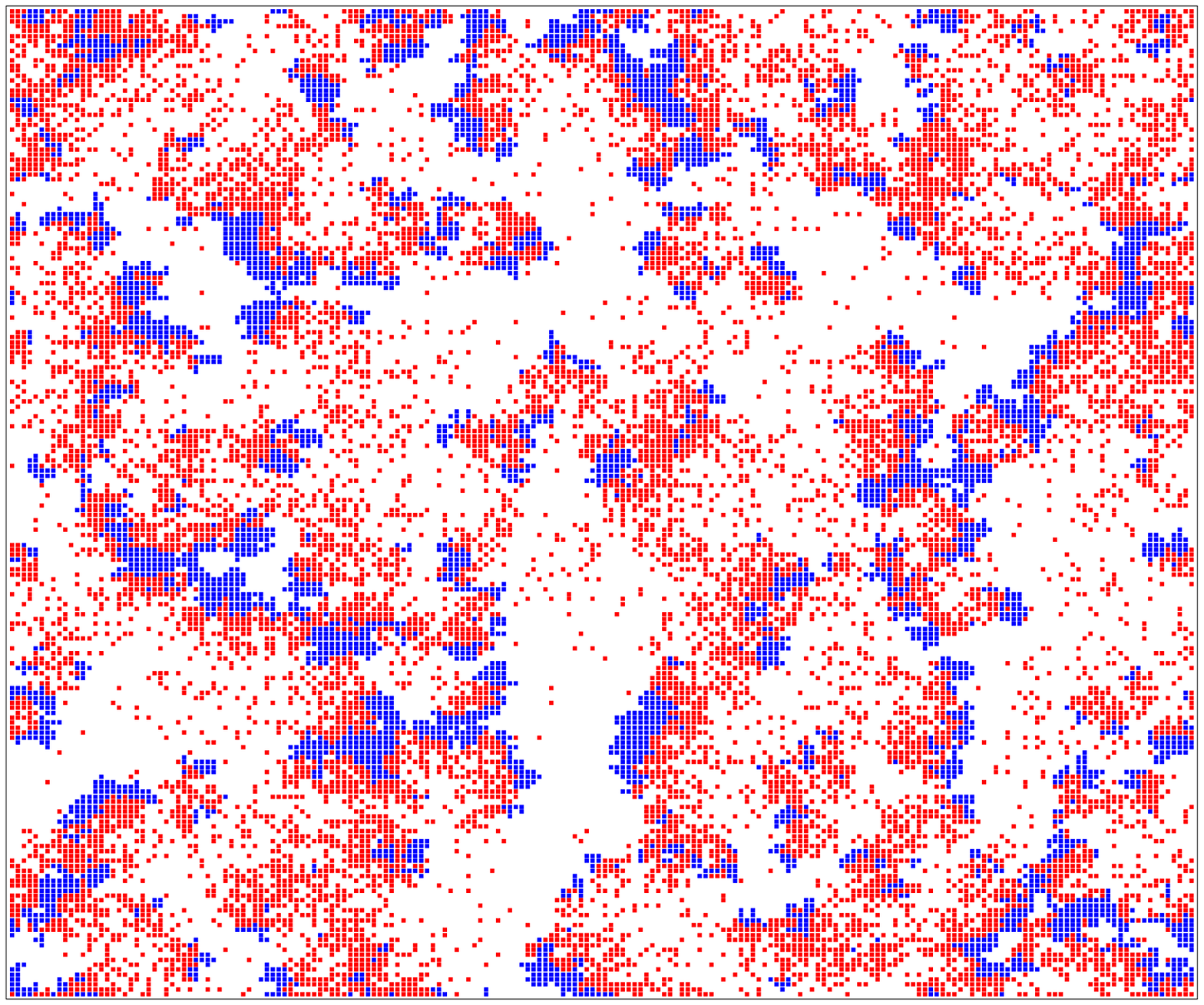,angle=-90,width=6.5cm}
\epsfig{file=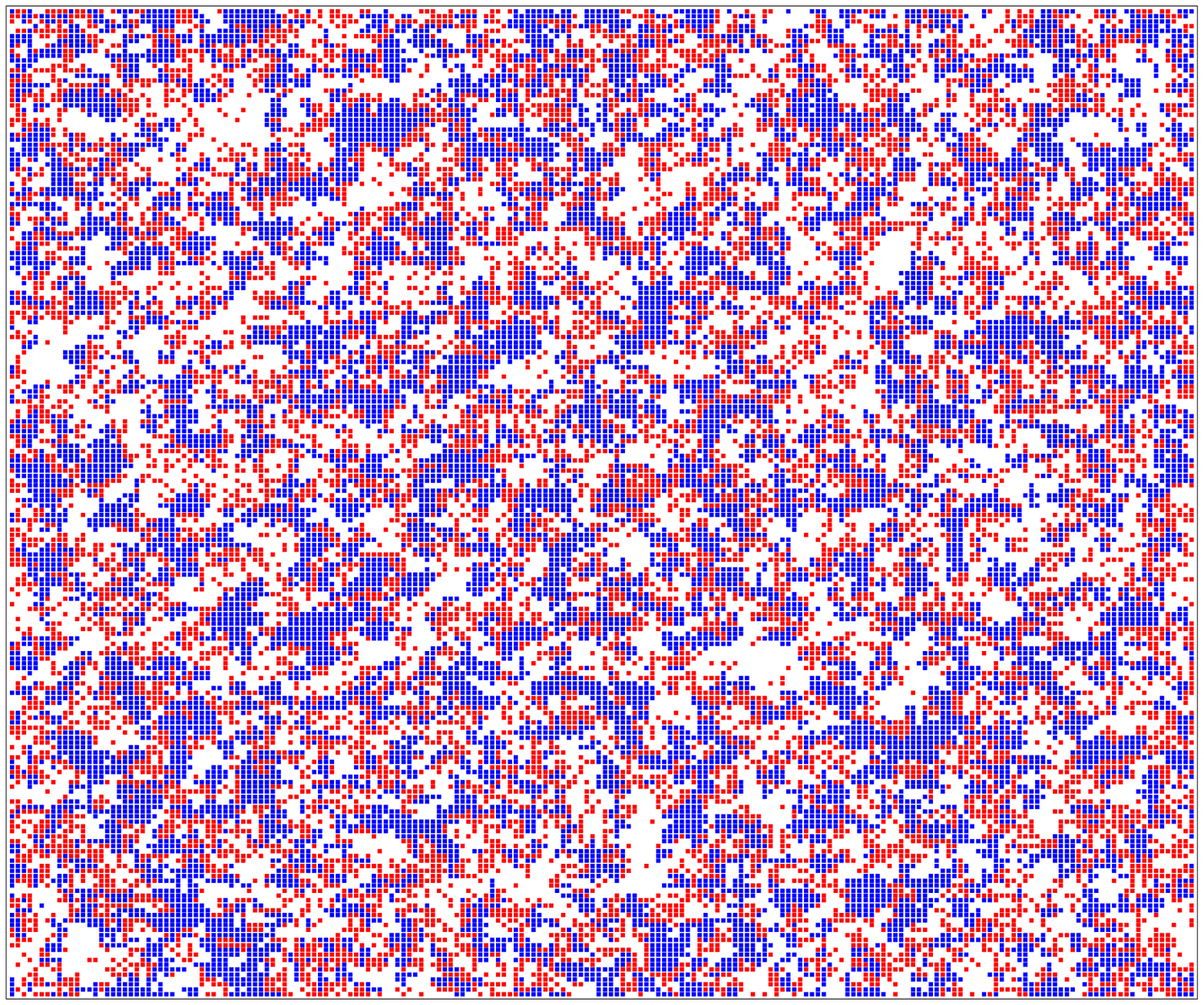,angle=-90,width=6.5cm}
\caption{Snapshots of the lattice (inside the oscillating region
of the $p-c$ diagram)
for $p=0$ and $c=0.03$ (left) and $p=0$ and  $c=0.10$ (right).
Results obtained for a lattice of size $160\times 160$.
The blue points represent sites occupied by prey, the red points 
by predators and the white points are empty sites.}
\label{mancha1}
\end{figure}

{\it Many cluster landscape oscillations}. 
For $p>0$ ($a>b$), typical spatial patterns
connected to the time oscillations can be seen in Figure \ref{mancha}.
We have analyzed the snapshots of the lattice taken at
successive instants of time.
In this figure, just the snapshots 
corresponding to the maximum of prey and 
to the minimum of prey population are shown. 
It can be observed that they are very
similar, and resemble a space covered with patches
occupied by the different species.
It is noticeable the presence of clusters of prey of different sizes
including a very large one, almost percolating
the lattice, when the maximum prey population is
attained; after this instant of time, the largest clusters
begin to breakdown into small clusters or to have their size reduced,
giving room to the predators which then attain their
maximum population. From this point on the prey population decreases until it
attains its minimum value. Large clusters of empty sites
occupy the lattice and predators start dying until they reach
their minimum population value. After this, prey begin to reproduce
until they reach a maximum population value closing a cycle
of coupled oscillation.
We remark that, in each cycle, the maximum of predator  
follows the maximum of prey.

In the region of the $p-c$ diagram
where $p\leq 0$ ($a\geq b$) and for small values of $c$,
the model can exhibit the patterns of coexistence of
species shown in Figure \ref{mancha1}. 
For sufficient low values of $c$ the density of predators is appreciable. 
Predators stay grouped together in small clusters distributed over the
entire lattice. The prey reproduce with rate
greater than (or equal to) their death rate. Any decrease
in prey population by predation is then quickly recovered. 
As a consequence the amplitudes of oscillations become small,
as seen in Figure \ref{osc}.
For a not so small value of $c$, as that corresponding to
the pattern shown in the right panel of Figure \ref{mancha1},
the densities of prey and predator still oscillate in time
but now the prey population in greater than the predator
population.

\begin{figure}
\centering
\epsfig{file=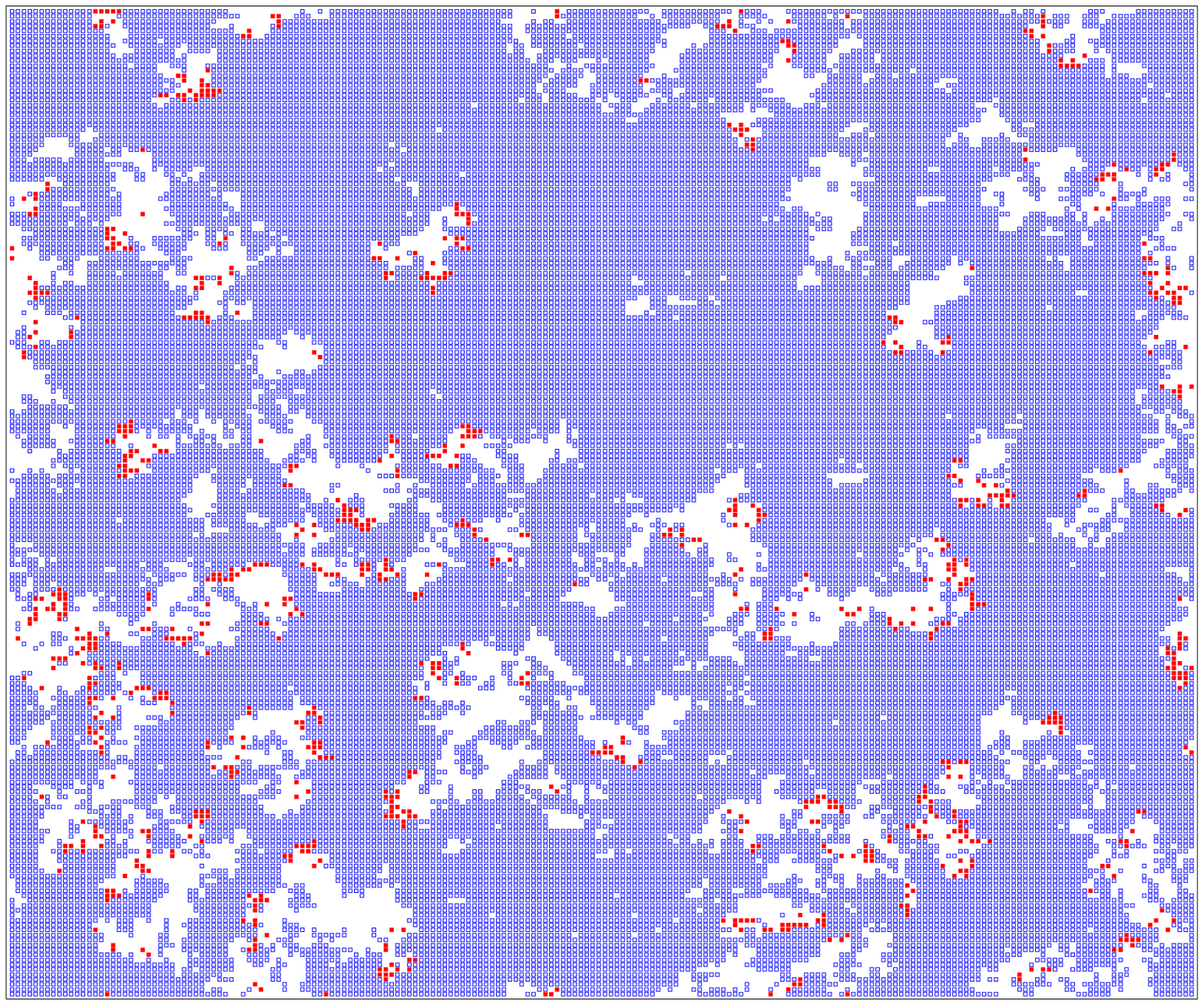,angle=-90,width=6.5cm}
\epsfig{file=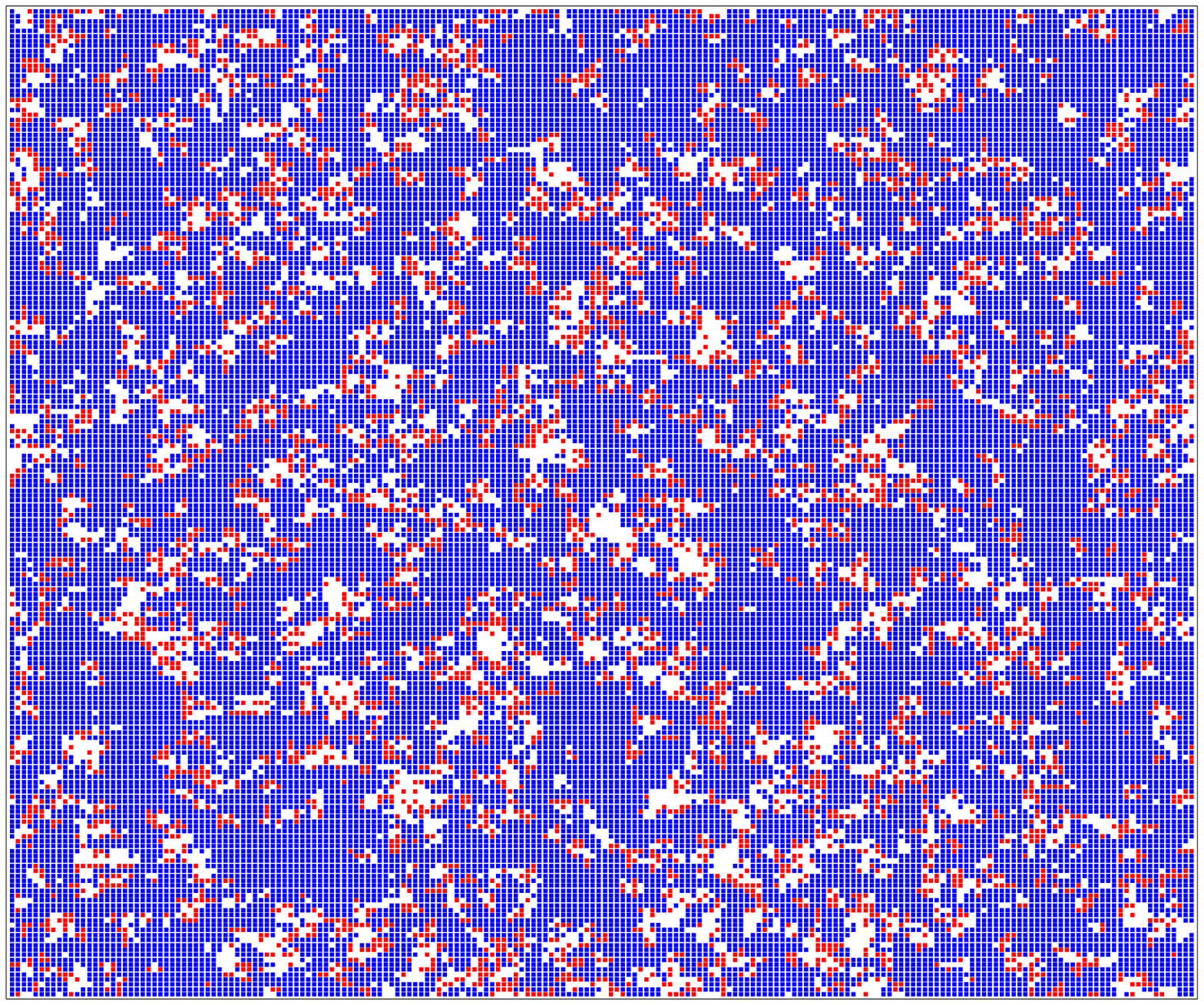,angle=-90,width=6.5cm}
\caption{Snapshots of the lattice for $p=0.3$ and $c=0.21$ (left)
and $p=0$ and $c=0.17$ (right), both inside the nonoscillating
region of the $p-c$ diagram. 
Results obtained for a lattice of size $160\times 160$.
The blue points
represent sites occupied by prey, the red points 
by predators and the white points are empty sites.
In both cases the blue points percolate the lattice.
}
\label{percol}
\end{figure}

{\it Prey percolating landscape}. 
For a larger value of $c$ the active nonoscillating region 
of the $p-c$ diagram is reached.
Snapshots of the lattice, showing patterns corresponding to
two points in the nonoscillating region of the $p-c$ diagram, 
are presented in Figure \ref{percol}. 
They correspond to the same $p$ values of Figures \ref{mancha} and 
\ref{mancha1}, but for a greater value of $c$. 
In these cases, there are a large cluster of prey that 
percolates the lattice and small clusters of empty sites
and of predators. 
Let us consider the case $p=0.3$ and compare the spatial
patterns at $c=0.13$, associated to the active oscillating state
(Figure \ref{mancha}), with the
pattern at $c=0.21$, associated to the active nonoscillating state
(left panel of Figure \ref{percol}). We observe one important feature that 
differentiates the two kind of active states for a fixed value of $p$.

In one case (active states without oscillations) 
there is a percolating cluster of prey and
in the other (active states with oscillations)
spatial patterns do not present a percolating cluster.
Similar comparison can be made for $p=0$
(see right panels of both Figures \ref{mancha1} and \ref{percol}). 
Therefore the onset of the nonoscillating active state
seems to be associated to the formation of
a percolating cluster of prey.
Presently, we are analyzing the
transition line in the $p-c$ diagram 
from the active nonoscillating state to the active
oscillating state, and its relation with the pattern formation and 
the percolation theory.

\begin{figure}
\centering
\epsfig{file=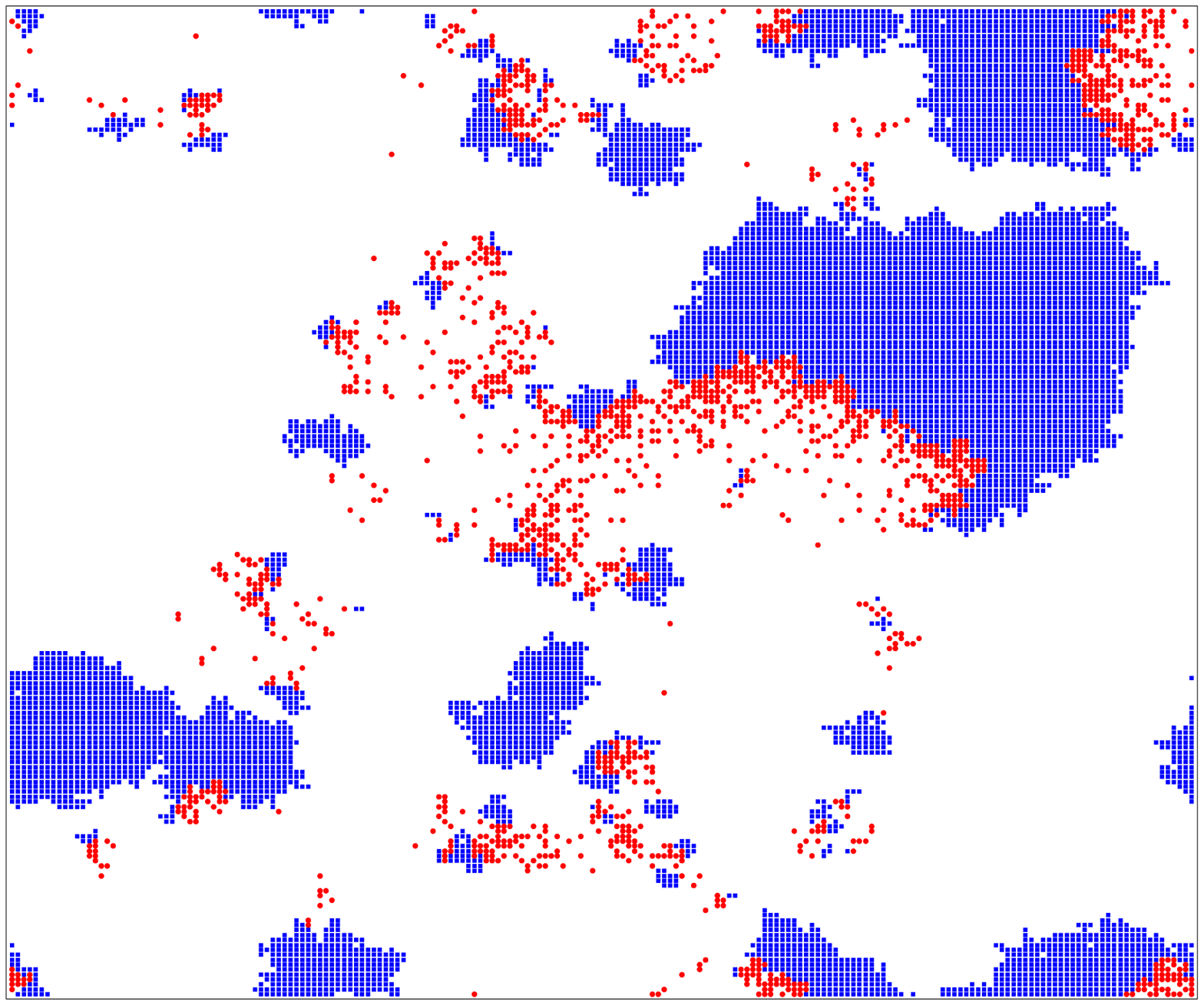,angle=-90,width=6.5cm}
\epsfig{file=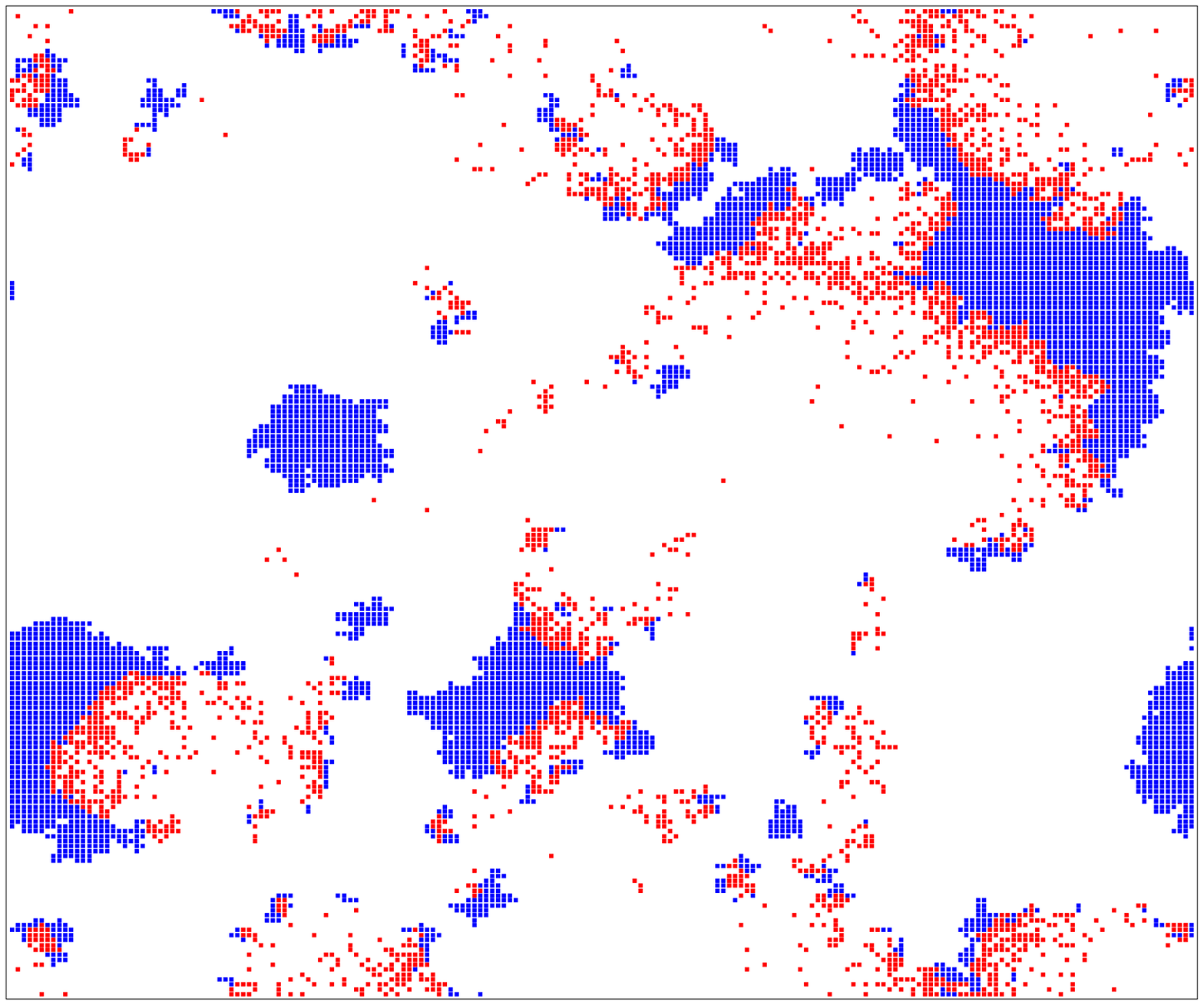,angle=-90,width=6.5cm}
\\[1mm]
\epsfig{file=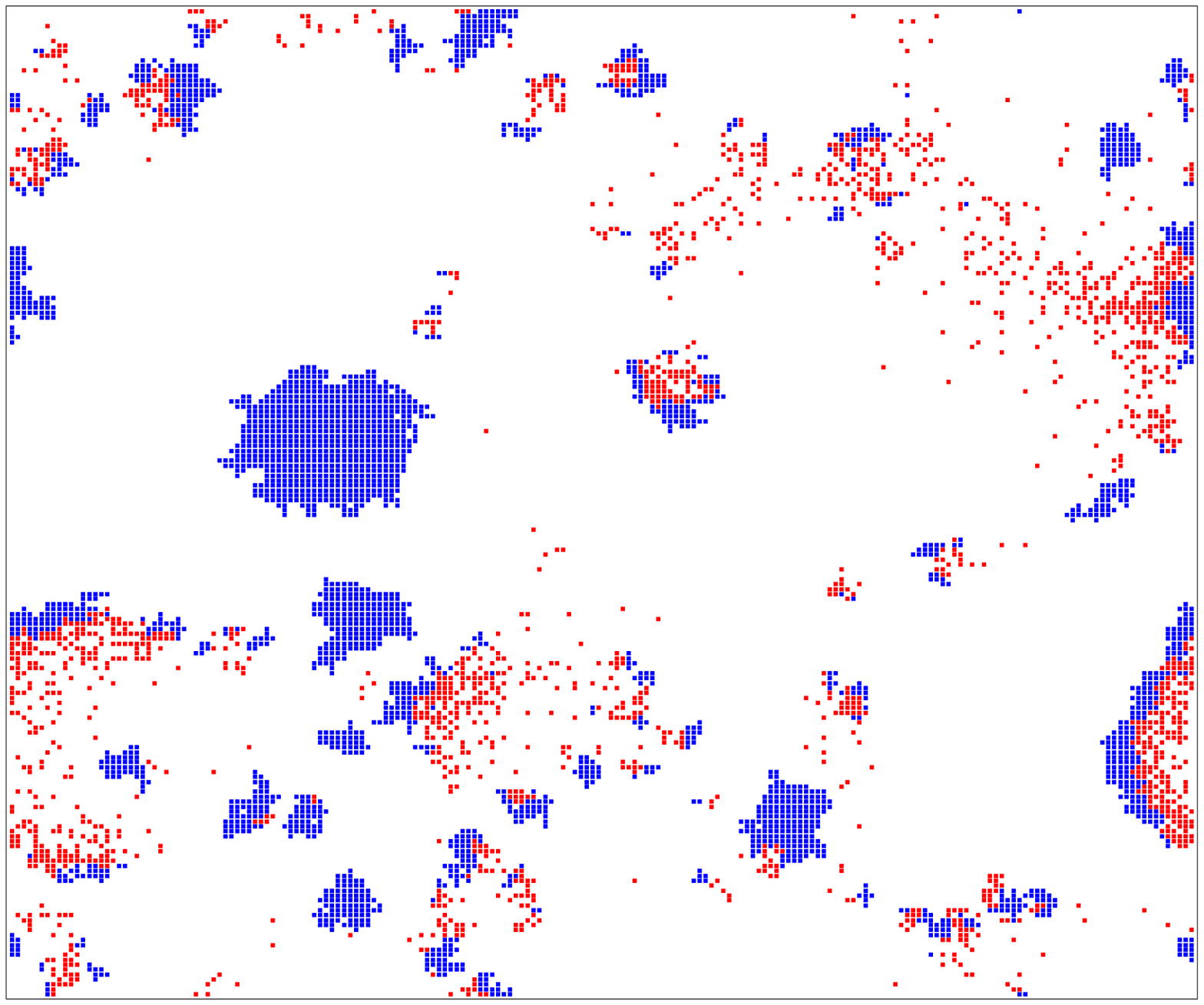,angle=-90,width=6.5cm}
\epsfig{file=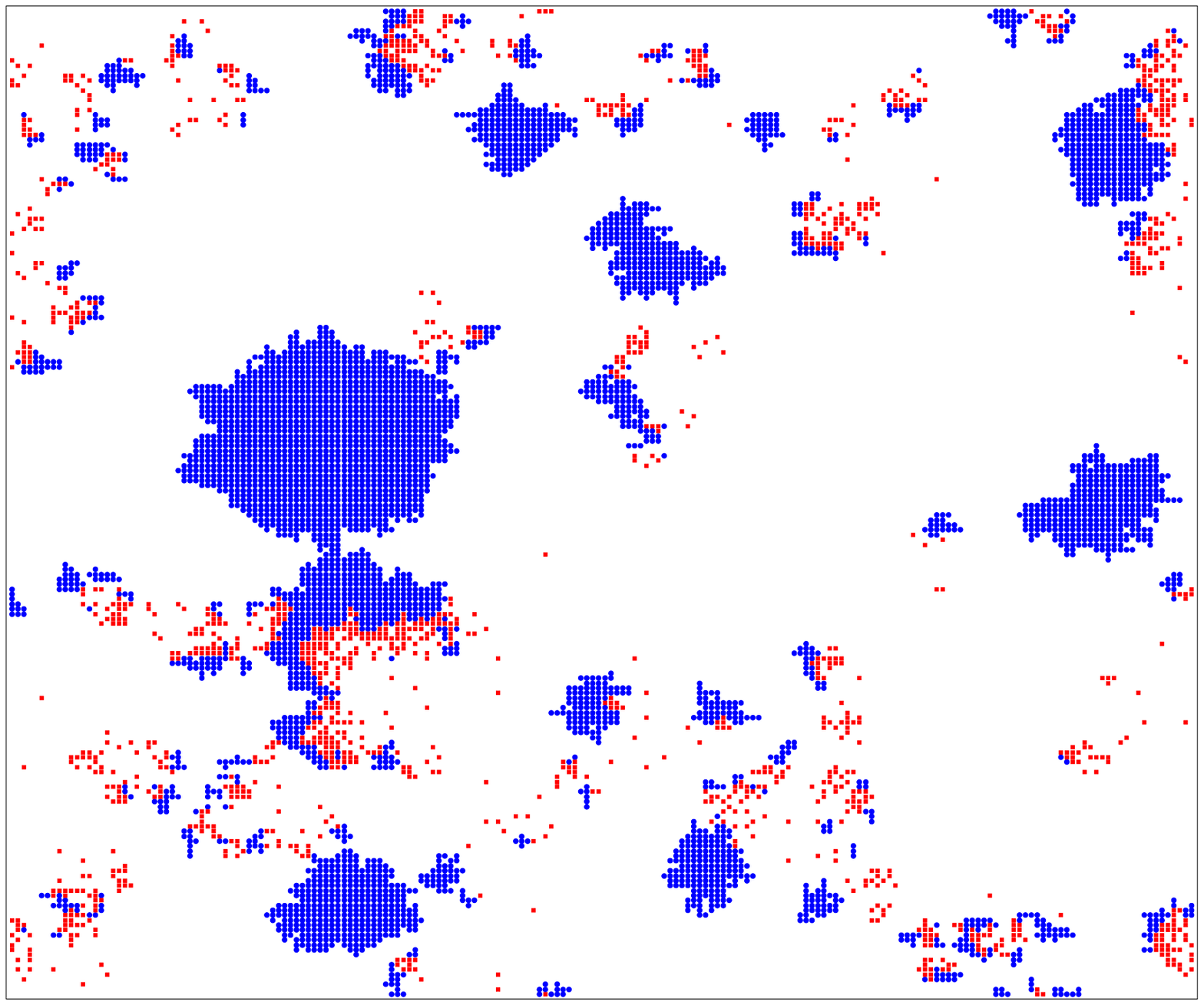,angle=-90,width=6.5cm}
\caption{Snapshots of the lattice for $p=0.3$ and $c=0.06$
(inside the oscillating region of the $p-c$ diagram) 
taken at maximum of prey (top, left),
maximum of predators (top, right),
minimum of prey (bottom, left), and 
minimum of predator (bottom, right).
Results obtained for a lattice of size $160\times 160$.
The blue points
represent sites occupied by prey, the red points 
by predators and the white points are empty sites.}
\label{ciclo}
\end{figure}

{\it Compact-cluster landscape oscillations}.
For $p>0$ ($a < b$)  and considering very small values of $c$,
we can observe a special spatial pattern formation
connected to the oscillations,
as seen in Figure \ref{ciclo}.
We are considering $p=0.3$, what implies 
that the death of prey occurs with probability
greater than its reproduction probability.
This can lead to a situation where a very small number
of prey is present on the lattice (minimum of prey).
On the other hand, under this situation, predators can 
reduce dramatically their reproduction, and even if their
probability of death $c$ is small, the predator density can evolve
to a very small value (minimum of predators). 
This fact allows a great increase of the isolated clusters of prey
until they become large and compact clusters (maximum of prey).
These compact clusters keep increasing until they
eventually encounter the reminiscent predators
or small groups of predators. At this moment the
predators eat prey very quickly, reproduce at a 
high rate, and attain their maximum population.
This is an example of the type of a sequence of pattern formation
connected to the time oscillations,
for set of the parameters, where the
predation is highly efficient, the birth prey probability is small
and the mean lifetime of predators is high ($c<<1$);
under these conditions 
predators are able to practically decimate a large cluster of prey.
And then, without food, they start dying, until the situation where 
small clusters of prey begin to increase and the
all succession of above described situations repeat in time
with a given characteristic frequency.
We observe that a maximum of
predator population  always follow a maximum of prey population.

\begin{figure}
\centering
\epsfig{file=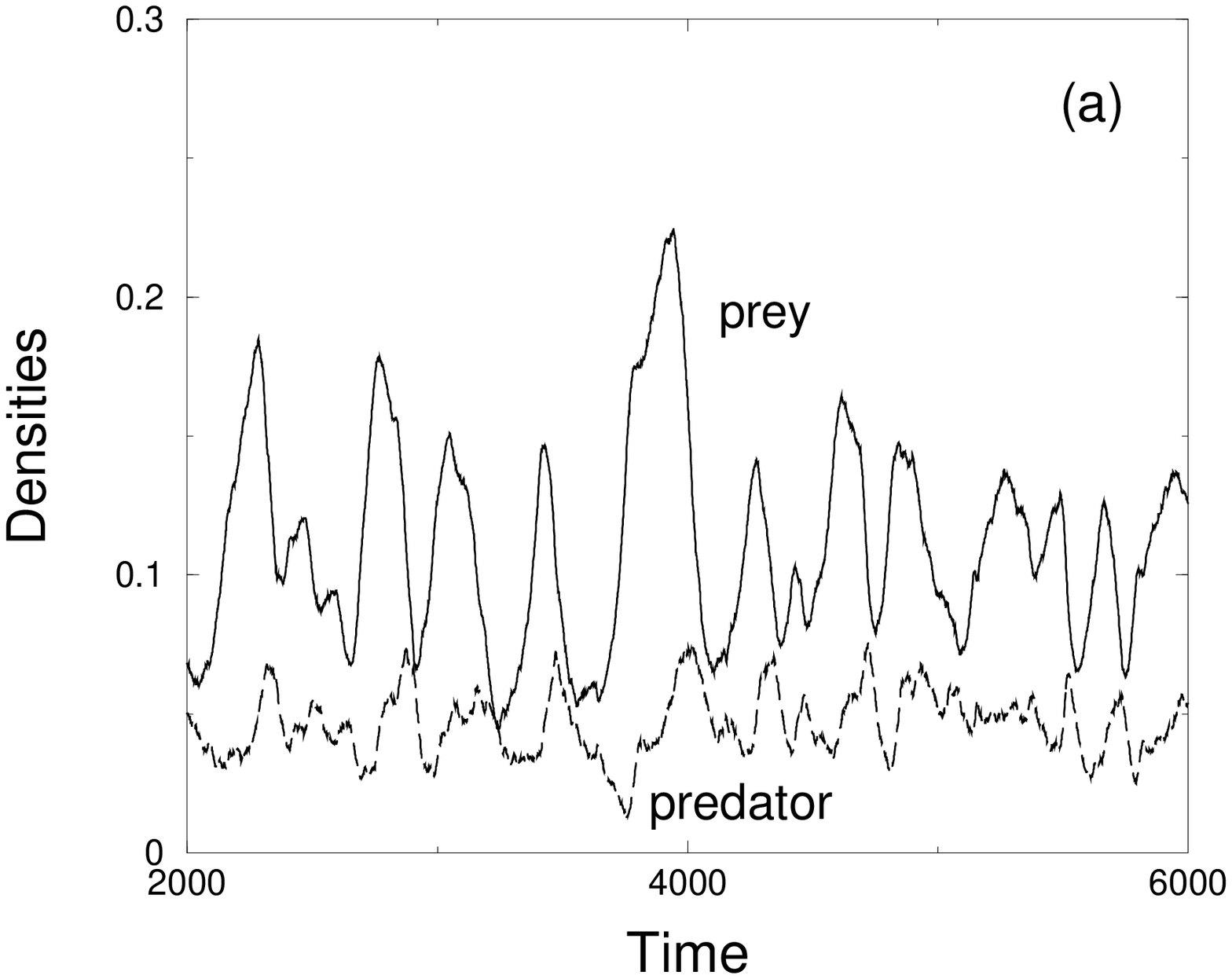,width=6.5cm}
\hfill
\epsfig{file=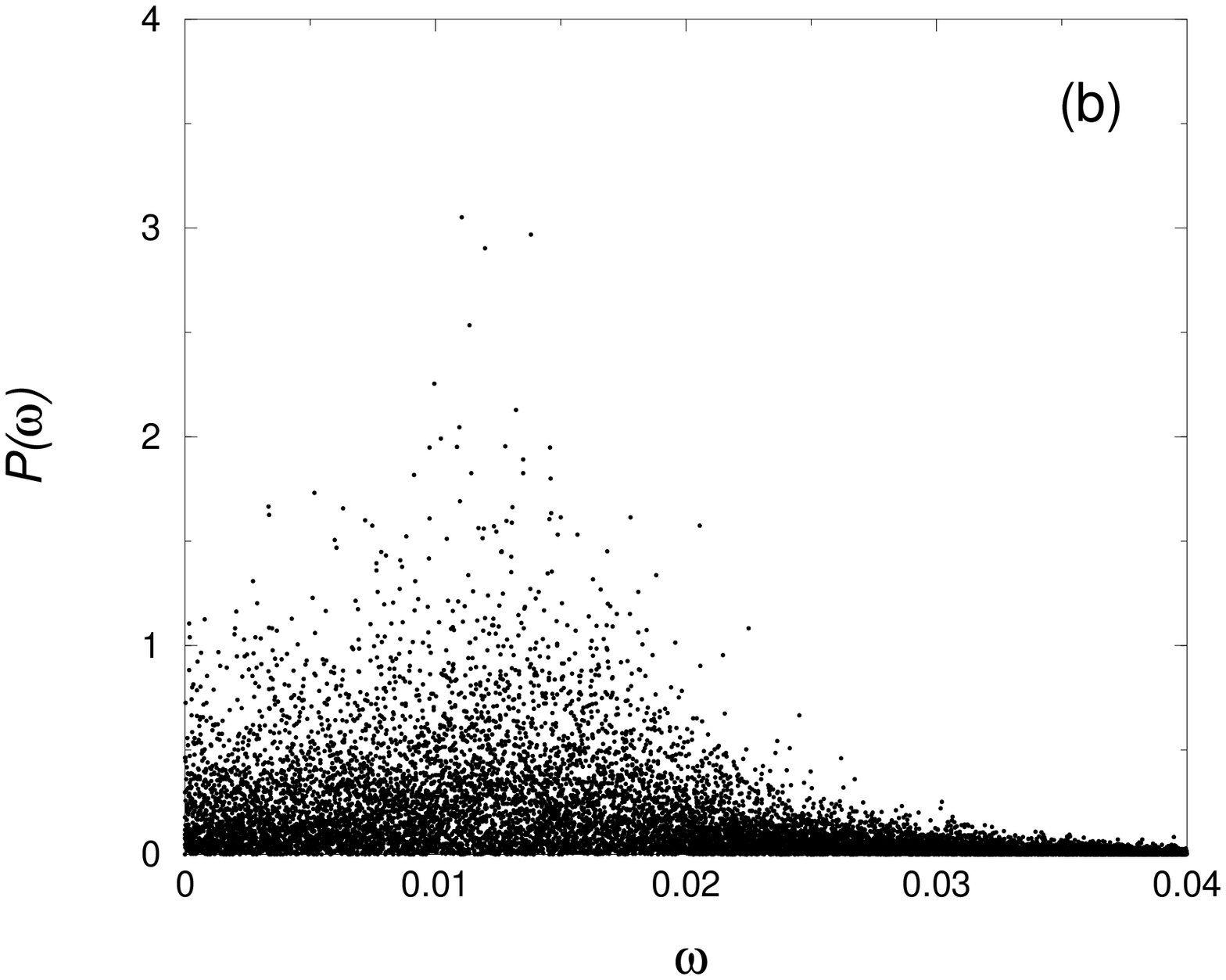,width=6.5cm}
\caption{(a) Densities of prey and predator as a function of 
time for $p=0.3$ and $c=0.06$
(inside the oscillating region of the $p-c$ diagram). 
(b) Corresponding power spectrum
for the density of prey as a function of the frequency $\omega$.
Results obtained for a lattice of size $160\times 160$.}
\label{suposc}
\end{figure}

The above mechanism leads
to most pronounceable amplitudes of oscillations,
as can be seen in Figure \ref{suposc}.
In this case the period of a cycle is large when compared
with the period of a cycle associated to the
many cluster landscape oscillations.
The large period is the result of a very slow increase in
the number of prey ($a<<b$) which might reach very large values
before starting to be decimated by the small number of
remaining predator individuals.

All the pattern formation, described above, are self-organized
structures (Nicolis and Prigogine, 1977; Haken, 1983; 
Tom\'e and de Oliveira, 1989; Hassel et al., 1994)
resulting from the spatio-temporal dynamics of the probabilistic
cellular automaton. It is important to observe that any macroscopic
ordering coming from a microscopic irreversible dynamics is
called a self-organized structure. Of course, here, this
phenomenon is more evident in the case of the highly
inhomogeneous spatial patterns of figure \ref{ciclo},
which are coupled to the self-sustained oscillation shown in
figure \ref{suposc}.

\section{Summary and discussion}

We have presented a probabilistic cellular automata with Markovian local
rules mimicking the interactions of a predator-prey system. In this
description we have considered a regular square lattice where each
site can be either occupied by at most one individual of each species or
can be empty and  the interactions just occur in the neighborhood of
a site. Our computational simulations show
that the model displays three states, depending on the 
values of the set of parameters:
the absorbing prey state, an active state where both
species coexist and their densities are constant in time
(nonoscillating active state) and an active state where a
self-sustained time oscillation of the prey and predator populations
is present.

The time oscillations of prey and predator populations are
related to spatial patterns characterized by clusters of prey and
predators. In this case, the prey clusters may cover large
regions of space but never percolate the lattice. We classified
the patterns associated to oscillation in two types. One of them,
which was called many cluster landscape, is constituted by a
large number of clusters distributed over the entire lattice.
The clusters are not compact and may be of a fractal nature.
In this case the clusters of prey grow and shrink
as the system attains a maximum or a minimum of prey
population in a cycle of oscillation. This spatio-temporal 
behavior is accomplished in a large range of the parameters.
The other type of pattern associated to oscillations,
which was called compact-cluster landscape oscillations,
is characterized by a few number of large compact clusters of prey.
In this case, the spatial patterns are very inhomogeneous and
they differ among themselves, apreciably, as the maximum
and minimum of each species is attained in a cycle. These
inhomogeneous patterns occur for small values of the prey 
birth probability and for very small values of death predator probability
(when compared to the predation probability values).
Under these conditions predators eat prey
so efficiently that they practically decimate all prey clusters 
becoming then isolated 
from the few and small reminiscent cluster of prey.
Without neighboring prey, they start to die 
leaving large regions of empty sites; a propitious condition for
the growing of the big and compact clusters of prey.    
Great periods and amplitudes of oscillations are detected.
The maximum of predators always follow the abundance of prey
in each cycle of oscillations

Our model also predicts a coexistence without oscillations,
which was called prey percolation landscape.
This regime happens on a small range of the parameters,
at values of predator death probability large compared with the
ones associated to the oscillating region. In this
case the number of prey is large and
the spatial pattern is characterized by the presence
of a percolating cluster of prey, that is, the 
prey get together in a extensive and connected region of  
space.
 
The results of the predator-prey cellular automaton indicate that the
spatio-temporal patterns of coexistence arises from a combination of
the Lotka-Volterra basic mechanisms of interaction, the discreteness
of individuals and the spatial-structured model. 
The self-sustained coupled oscillations of prey and predator
populations appear, in finite systems, when the death probability of
predators is sufficiently small. The coexistence without oscillations
appear when this probability is increased. The threshold of the
transition between the regime of coexistence of species and the regime 
where predators have been extincted is enhanced as the predation
probability becomes greater than the birth prey probability.
For very small values of the predator death probability and high values
of the predation probability the spatial patterns associated to
oscillations are highly inhomogeneous. The pattern formation, as well
as the time oscillations, result from the intrinsic dynamics of the
probabilistic cellular automaton. In this sense, they are self-organized
structures, that is, the collective interactions of the system of
particles in the lattice is able to produces spatio-temporal
macroscopic ordering which is self-maintained.

\section*{Acknowledgments}

This work was partially supported by the Brazilian agency 
Conselho Nacional de Pesquisa (CNPq).

\section*{References}

\small

\begin{trivlist}

\item[\sc Aguiar M.A.M., Rauch E.M., and Bar-Yam Y., 2003.]
Mean-field approximation to a spatial host-pathogen model,
Phys. Rev. E 67, 047102.

\item[\sc Antal T., Droz M., 2001.]
Phase transitions and oscillations in a lattice prey-predator model,
Phys. Rev. E 63, 056119.

\item[\sc Carvalho K.C., Tom\'e T., 2004.]
Probabilistic cellular automaton describing a 
biological two-species system,
Mod. Phys. Lett. B 18, 873-880.

\item[\sc Caswell H., Etter R.J., 1993.]
Ecological interactions in patch environments:
from patch-occupancy models to cellular automata, 
in Levin S.A., Powell T.M. (Eds.),
Patch Dynamics, Springer, New York, pp 93-109,

\item[\sc Durrett R. 1988.] 
Lecture Notes on Particle Systems and Percolation, 
Wadsworth and Brooks, Pacific Grove, California.

\item[\sc Durrett R. and Levin S., 1994.] 
The importance of being discrete (and spatial),
Theor. Popul. Biol. 46, 363-394.

\item[\sc Durrett R. and Levin S., 2000.] 
Lessons on patern formation from planet WATOR,
J. Theor. Biol. 205, 201-214.

\item[\sc Grassberger P., 1982.]
On phase transitions in Schogl's second model,
Z. Phys. B 47, 365-374.

\item[\sc Haken H., 1983.]
Synergetics: An Introduction, Springer, Berlin.

\item[\sc Hanski I., Gilpin M.E., (Eds.), 1997.]
Metapopulation Biology: Ecology, Genetics and Evolution,
Academic Press, New York, pp. 123-146.

\item[\sc Hassel M.P., Comins H.N., May R.M., 1994.]
Species coexistence and self-organizing spatial dynamics,
Nature 370, 290-292.

\item[\sc Hastings A., 1997.]
Population Biology: Concepts and Models,
Springer, New York.

\item[\sc Huffaker C.B., 1958.] 
Experimental studies on
predation: dispersion factors and predator-prey oscillations,
Hilgardia 27, 343-383.

\item[\sc King A.A., Hastings A., 2003.]
Spatial mechanisms for coexistence of species sharing a
common natural enemy,
Theor. Popul. Biol. 64, 431-438. 

\item[\sc Levin, S.A., 1974.]
Dispersion and population interactions, 
Am. Naturalist 108, 207-228.

\item[\sc Liggett T.M., 1985.]
Interacting Particle Systems, Springer, New York.

\item[\sc Liu Y.C., Durrett R., Milgroom M., 2000.]
A spatially-structured sto\-cha\-stic \break model to simulate
heterogeneous transmission of viruses in fungal populations,
Ecol. Model. 127, 291-308.

\item[\sc Lotka A., 1920.]
Analytical note on certain rhythmic relations in organic systems,
Proc. Nat. Acad. Sci. 6, 410-415.

\item[\sc Marro J., Dickman R., 1999.]
Nonequilibrium Phase
Transitions in Lattice models, Cambridge University Press,
Cambridge.

\item[\sc Nakagiri N., Tainaka K., Yoshimura J., 2005.]
Bond and site percolation and habitat
destruction in model ecosystems, J. Phys. Soc. Jpn. 74, 3163-3166.

\item[\sc Nicolis G., Prigogine I., 1977.]
Self-Organization in Nonequilibrium Systems,
Wiley, New York.

\item[\sc Ovaskainen O., Sato K., Bascompe J., Hanski I., 2002.]
Metapopulation models for extinction threshold
in spatially correlated landscapes,
J. Theor. Biol. 215, 95-108.

\item[\sc Provata A., Nicolis G., Baras F., 1999.]
Oscillatory dynamics in low-dimensio\-nal supports: a lattice
Lotka-Volterra model, J. Chem. Phys. 110, 8361-8368.

\item[\sc Satulovsky J.E., Tom\'e T., 1994.]
Stochastic lattice gas model for a predator-prey system,
Phys. Rev. E 49, 5073-5079.

\item[\sc Satulovsky J. E., Tom\'e T., 1997.]
Spatial instabilities and local oscillations in a
lattice gas Lotka-Volterra model,
J. Math. Biol. 35, 344-358.

\item[\sc Stauffer D., Kunwar A., Chowdhury D., 2005.]
Evolutionary ecology in silico:
evolving food webs, migrating population and speciation,
Physica A 352, 202-215.

\item[\sc Szab\'o G., Sznaider G. A., 2004.]
Phase transition and selection in a four-species cyclic 
predator-prey model, Phys. Rev. E 69, 031911.

\item[\sc Tainaka K., 1988.]
Lattice model for the Lotka-Volterra system,
J. Phys. Soc. Japan 57, 2588-2590.

\item[\sc Tilman D., Kareiva P., 1997.]
Spatial Ecology: The R\^ole
of Space in Population Dynamics and Interactions,
Princeton University Press, Princeton.

\item[\sc Tom\'e T., 1994.]
Spreading of damage in the Domany-Kinzel cellular automaton:
a mean-field approach, Physica A 212, 99-109.

\item[\sc Tom\'e T., de Oliveira M.J., 1989.]
Self-organization in a kinetic Ising model,
Phys. Rev. A 40, 6643-6646.

\item[\sc Tom\'e T., de Oliveira M.J., 2001.]
Din\^amica Estoc\'astica e Irreversibilidade,
Edi\-to\-ra da Universidade de S\~ao Paulo, S\~ao Paulo
(in Portuguese).
 
\item[\sc Tom\'e T., Drugowich de Felicio J.R., 1996.]
Probabilistic cellular automaton describing a biological immune system,
Phys. Rev. E 53, 3976-3981.

\item[\sc van Kampen N.G., 1981.]
Stochastic processes in Physics and Chemistry,
North-Holland, Amsterdam.

\item[\sc Volterra V., 1931.] 
Le\c{c}on sur la Th\'eorie Math\'ematique de la Lutte pour la Vie,
Gauthier-Villars, Paris.

\end{trivlist}

\end{document}